\documentclass[11pt,twoside]{article}

\usepackage{asp2004}
\usepackage{epsf}
\usepackage{psfig}
\usepackage{lscape}
\def\kms{km~s$^{-1}$}
\def\kms{km~s$^{-1}$}
\begin{document}

\title{Properties of the thinnest cold HI clouds in the diffuse 
interstellar medium}   
%%%Fill in title

\author{Sne\v{z}ana Stanimirovi\'{c}}   %%% Fill in author names
\affil{Radio Astronomy Lab, UC Berkeley, 601 Campbell Hall,
Berkeley, CA 94720; Present address: Department of Astronomy, University of Wisconsin, 
475 North Charter Street, Madison, WI 53706}
\author{Carl Heiles}
\affil{Radio Astronomy Lab, UC Berkeley, 601 Campbell Hall,
Berkeley, CA 94720}
\author{Nissim Kanekar}
\affil{National Radio Astronomy Observatory, 1003 Lopezville Road, 
Socorro, NM 87801}

\begin{abstract} %%% Abstract to run on from here.
We have obtained deep HI observations in the direction of 22 continuum sources
without previously detected cold neutral medium (CNM). 18 CNM 
clouds were detected with the typical HI column density of 
$3\times10^{18}$ cm$^{-2}$. 
Our surprisingly high detection rate suggests 
that clouds with low HI column densities are quite common in the interstellar
medium. These clouds appear to represent an extension of the
traditional CNM cloud population, yet have sizes in hundreds to thousands of
AUs. We present properties of the newly-detected CNM sample, and
discuss several theoretical avenues important for 
understanding the production mechanisms of these clouds. 
\end{abstract}

%%% MAIN BODY OF TEXT GOES HERE. CONSULT "INSTRUCTIONS FOR AUTHORS USING
%%% LATEX2E MARKUP", SECTIONS 2.3-2.6 FOR HELP WITH EQUATIONS, FIGURES,
%%% AND TABLES.

\section{Introduction}   %%% Top level section head (remove "%" symbol)

While properties and origin of the AU-scale structure in the 
cold neutral medium (CNM) are still under debate, 
a possibly related new population of CNM clouds 
has been dicovered recently. Using the Westerbork radio 
telescope, Braun \& Kanekar (2005) detected
very weak HI absorption lines toward three high-latitude sources.
Along each line of sight multiple absorption lines were detected, 
with the peak optical depth of only 0.1 to 2\%.
Stanimirovic \& Heiles (2005) used the Arecibo telescope 
to confirme the existence of these low-optical-depth CNM clouds 
in directions of two of the sources. 
They also emphasized that these clouds have HI column densities 
among the lowest ever detected for the CNM, $\sim10^{18}$ cm$^{-2}$.
We will therefore call these clouds 
the `low-$N$(HI)' clouds.

\begin{figure}[!ht]
\plotfiddle{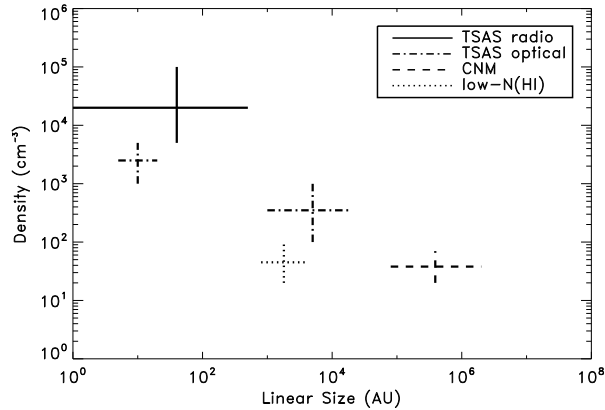}{5cm}{0}{50}{50}{-170}{-115}
\caption{\label{f:size_density} Typical linear size and HI volume density of
  TSAS, traditional CNM clouds (dashed line), and the low-$N$(HI) clouds 
(dotted line; based on data from Stanimirovic \& Heiles 2005). 
In the case of TSAS, the radio measurements are
shown with a solid line (from Heiles, SINS), while optical measurements
  of double stars and stars in globular and open clusters 
(probing scales of  hundreds of AU), as well as 
  from temporal variability (probing scales of tens of AU), 
are shown with dot-dashed lines (from Lauroesch, SINS).}
\end{figure}

How atypical are low-$N$(HI) clouds?
From the theoretical point of view, the traditional CNM clouds have a typical
size of 2 pc and $N({\rm HI})\sim3\times10^{20}$ cm$^{-2}$, 
the lowest expected column density being $6\times10^{19}$ cm$^{-2}$ (McKee \&
Ostriker 1977). From an observational point of view, the recent
survey by Heiles \& Troland (2003, HT03) suggested a typical 
$N({\rm HI})=0.5\times10^{20}$ cm$^{-2}$ for CNM clouds.
While column densities of low-$N$(HI) clouds are
30--50 times lower than theoretical and observational
expectations, these densities are close to what is 
measured for the tiny scale atomic structure (TSAS), 
$N({\rm HI})\sim {\rm a~few} \times 10^{18}$ to $10^{19}$ cm$^{-2}$ 
(Heiles, SINS).
In Figure~\ref{f:size_density} we illustrate graphically 
how low-$N$(HI) clouds compare with TSAS and CNM clouds
by plotting the typical linear size and HI volume density for these three
types of objects.
Low-$N$(HI) clouds occupy the region in this diagram between
TSAS and CNM clouds, the regime that is currently observationally probed 
only with optical observations of globular clusters 
(e.g. Meyer \& Lauroesch 1993).

In this contribution we focus on two particular questions regarding the
low-$N$(HI) clouds:
(1) how common are these clouds in the ISM, and
(2) how are these clouds related to the traditional spectrum of CNM clouds?
In Section~\ref{s:obs} we summarize our recent search for 
the low-$N$(HI) clouds with the Arecibo telescope. 
We describe our results in Section~\ref{s:results}
and discuss physical mechanisms responsible for the production of low-$N$(HI) 
clouds in Section~\ref{s:discussion}

\section{Recent search for low-N(HI) clouds with the Arecibo telescope}
\label{s:obs}

To search for new low-$N$(HI) clouds we have recently obtained HI emission and 
absorption spectra in the direction of 22 continuum sources with the Arecibo 
radio telescope. About half of the sources were chosen from 
HT03 as being without detectable 
CNM after $\sim15$ minutes of integration,  
the remaining sources were selected from catalogs by Dickey
et al. (1978) and Crovisier et al. (1980). 
None of the sources in our sample had previously detected CNM.
The observing strategy was the same as in HT03 and
Stanimirovic \& Heiles (2005), however
the integration time per source was significantly longer (1 to 4.5 hours). 
The final velocity resolution of HI spectra is 0.16 \kms. 
The final rms noise level in the absorption spectra 
is $\sim5\times10^{-4}$ over 0.5 \kms~channels. 
For sources with newly-detected CNM we used the technique developed 
by  HT03 to estimate the spin temperature.
However, this technique turns out to be unreliable for our data as
most of the CNM clouds have a very low optical depth and 
occupy a solid angle significantly smaller than the Arecibo beam. 
For CNM features presented in this paper we have chosen 
$T_{\rm sp} = T_{\rm k}/2$. This is a safe assumption that probably 
over-estimates our $N({\rm HI})$ as HT03 found that majority of 
their CNM clouds had $T_{\rm sp} = T_{\rm k}/3$.

\begin{figure}[!ht]
\plotfiddle{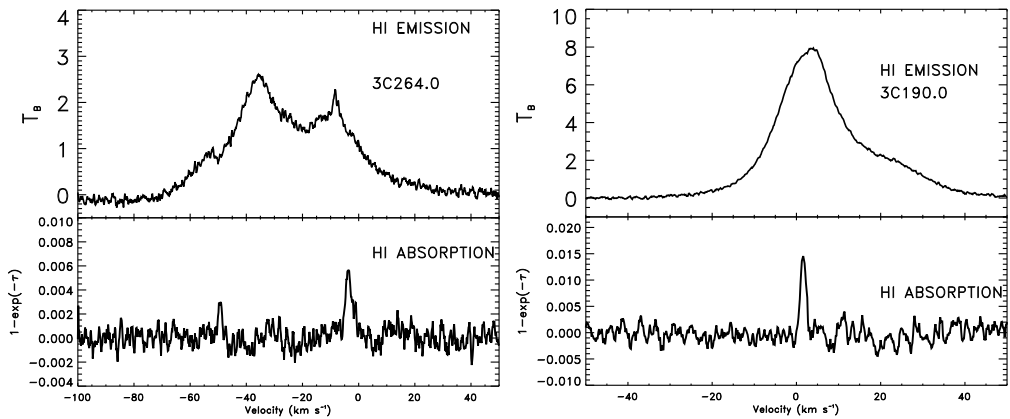}{5.cm}{0}{75}{75}{-230}{-240}
\vspace{0cm}
\caption{\label{f:3c}HI emission and absorption spectra for 3C264 (left) and
  3C190.0 (right) obtained with the Arecibo telescope. Two new CNM components
  were detected for 3C264, and one for 3C190.0.}
\end{figure}

\section{Properties of low-$N$(HI) clouds}
\label{s:results}

Out of 22 sources in this study 10 show clear CNM features, 
and in many cases multiple 
lines along the line of sight were detected.
In total, we have detected at least 18 new CNM features with 
the peak HI optical
depth in the range $2\times10^{-3}$ to $2\times10^{-2}$.
{\it The detection rate in this experiment is surprisingly high,
suggesting that clouds with low optical depth are quite
common in the ISM.} 

Figure~\ref{f:3c} shows HI emission and absorption spectra for two sources
in our survey, 3C264.0 (left) and 3C190.0 (right). 
For each source, the top and bottom panels show the HI emission and absorption
spectra.
We detected two CNM clouds in the case of 3C264.0 and one cloud in 
the case of 3C190.0.
The peak optical depth is $5\times10^{-3}$ and
$3\times10^{-3}$ for clouds in the direction of 3C264.0, 
and $2\times10^{-2}$ for the cloud in the direction of 3C190.0.
The velocity FWHM for the three clouds is 3.5, 1.5, and 1.4 \kms, respectively.
Clearly, these are cold HI clouds, 
with 
$N({\rm HI})\sim 5\times10^{18}$, $2\times10^{17}$, and
$1\times10^{18}$ cm$^{-2}$, respectively.  
The HI peak brightness temperature in these directions is only about 2.5 K and
$\sim8$ K, with the total $N$(HI) being $2\times10^{20}$ cm$^{-2}$ and
$3.3\times10^{20}$ cm$^{-2}$, respectively. The ratio of the CNM to total HI
column density, $R=N({\rm HI})_{\rm CNM}/N({\rm HI})_{\rm TOT}$ 
is only about 5\%
and $\sim1$\%, respectively.

In total, our current sample has 21 low-$N$(HI) clouds: 18 from this
study, and three from Stanimirovic \& Heiles (2005). 
Two thirds of the clouds have $N({\rm HI})<10^{19}$ cm$^{-2}$.
In comparison, HT03 had 20 clouds with $N({\rm HI})<10^{19}$ cm$^{-2}$ out of 
143 CNM components.
In comparison to the Millennium Survey by HT03, we have almost doubled the 
number of clouds in the lowest column density bin. % in HT05.
The median properties for the whole population are: $\tau_{\rm max}=10^{-2}$, 
FWHM=2.4 \kms, and $N({\rm HI})=3\times10^{18}$ cm$^{-2}$.

\begin{figure}[!ht]
\plotfiddle{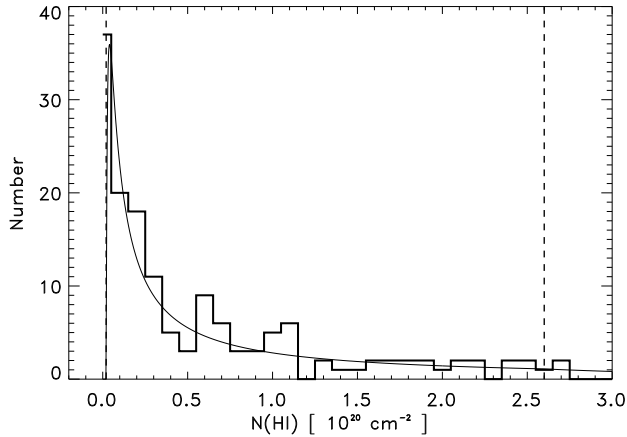}{5cm}{0}{45}{45}{-150}{-95}
\caption{\label{f:pdf}The observed distribution of $N({\rm HI})$ derived by
  combining the CNM components from HT03 with low-$N$(HI) clouds. 
The solid line represents the curve $N^{-1}$ defined over the 
range of column densities from $2\times10^{18}$
  to $2.6\times10^{20}$ cm$^{-2}$. The minimum and maximum column densities
  are shown with dashed lines.}
\end{figure}

The next obvious question to ask is how do low-$N$(HI) clouds relate to 
CNM clouds with higher column densities?
Heiles \& Troland (2005) investigated statistical properties 
of the CNM components detected in their Millennium Survey 
and found that the probability distribution of the CNM column density
closely follows $\Phi(N)\propto N^{-1}$ 
over two orders of magnitude, from $N_{\rm min}=2.6\times10^{18}$ cm$^{-2}$ to 
$N_{\rm max}=2.6\times10^{20}$ cm$^{-2}$.
We have added column densities of our 21 low-$N$(HI) clouds to $\Phi(N)$
from Heiles \& Troland (2005) and the resultant histogram is shown 
in Figure~\ref{f:pdf}.
The solid line in this figure represents 
the $\Phi(N)\propto N^{-1}$  fit obtained by slightly extending the lower
limit to $N_{\rm min}=2\times10^{18}$ cm$^{-2}$.
Surprisingly, a single function fits well both sets of objects.
This suggests that low-$N$(HI) clouds may not belong to a separate class 
of interstellar clouds, but could simply be {\it the 
low column density extension of the CNM population}.

In Figure 4 we show $R$ vs $N({\rm HI})_{\rm TOT}$ for all CNM components
found in the HT03 survey (shown with triangles and crosses) and in our
survey (circles). HT03 noticed that CNM clouds in their survey
followed almost a linear trend up to about 
$N({\rm HI})_{\rm TOT}\sim12\times10^{20}$ cm$^{-2}$, 
while sources without the CNM clearly stood out as
a potentially distinct class of objects.
Most low-$N$(HI) clouds have $R<5$\% which is suggestive of the CNM being 
embedded in large warmer envelopes that contain most of the HI and provide a
protection against evaporation.

If we assume that low-$N$(HI) clouds are at the standard ISM pressure of
$n({\rm HI})T \sim 3000$ K cm$^{-3}$, then their estimated  
HI volume density is $n({\rm HI})\sim 20$--100 cm$^{-3}$. 
The line-of-sight size of these clouds
is then given by $L(||) = N({\rm HI})/n({\rm HI})\sim 800$--4000 AU.
If the clouds are however overpressured
and $n({\rm HI})T > 3000$ K, then $L(||) < 800$--4000 AU, even
closer to the traditional TSAS.
In any case, low-$N$(HI) clouds are most likely small, 
with sizes in the range of a few
hundreds to a few thousands of AUs. This is larger than the `canonical' TSAS
but still significantly smaller than the traditional pc-size CNM clouds.
Another constraint on cloud sizes comes from direct interferometric
imaging by Braun \& Kanekar (2005) who reported HI emission clumps with 
sizes of $3\times10^{3}$ AU.
In any case, it appears that low-$N$(HI) clouds could be an
extension of the traditional population of CNM clouds, 
bridging the gap between CNM clouds and TSAS.

\section{Discussion}
\label{s:discussion}

Physical mechanisms responsible for the production of cold clouds with very
low column densities are not well understood.
There are three potentially interesting theortical avenues that await a
closer comparison with observations.

\begin{figure}[!ht]
\plotfiddle{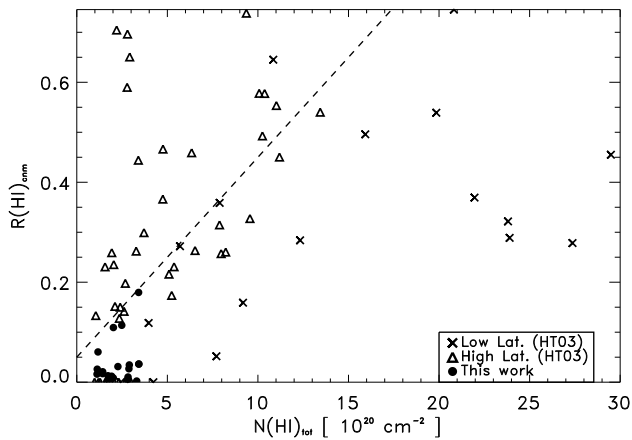}{5cm}{0}{45}{45}{-150}{-95}
\caption{CNM fraction $R=N({\rm HI})_{\rm CNM}/N({\rm HI})_{\rm TOT}$ vs 
$N({\rm HI})_{\rm TOT}$ for clouds from HT03, shown with crosses for latitudes
between 10 and 30 degrees and with diamonds for latitudes higher than 30
degrees. Low-$N$(HI) clouds are shown as filled circles. The dashed line
represents the trend noted by HT03.}
\end{figure}

{\it Conductive heat transfer} occurs between the CNM and their 
surrounding warmer medium, forming an interface region through 
which heat flows between the two phases. 
In the case of the least intrusive interface, CNM to WNM with the 
intercloud temperature of $10^{2}-10^{4}$ K, the expected lower limit on 
the column density of the interface region  is quite low, 
$N({\rm HI})\sim3\times10^{17}$ cm$^{-2}$ (McKee \& Cowie 1977). 
In this case the expected critical radius for CNM clouds 
(at which clouds are stable and neither evaporate nor condense) is 
$R_{\rm rad}\sim6\times10^{3}$ AU.
CNM clouds smaller than $R_{\rm rad}$ evaporate but 
are surprisingly long-lived, their typical evaporation timescale is
of order of $10^{6}$ yr (see Stanimirovic \& Heiles 2005 for details). 
Similar timescales were reached recently by Nagashima, Inutsuka, \& Koyama
(2006) based on 1-D numerical simulations.
Our low-$N$(HI) clouds have sizes below the critical cloud radius 
even in the case of the least-intrusive WNM. 
These CNM clouds, protected from ionization by
large WNM envelopes, should be evaporating but 
over long timescales, and hence could be common in the ISM.

 {\it Condensation of WNM into CNM triggered by the collision 
of turbulent flows} 
is capable of producing a large number of small CNM clouds with low column
densities, as seen in simulations by Audit \& Hennebelle (2005) and Hennebelle
(2006, SINS). The CNM clouds produced in simulations are
thermally stable and embedded in large, unstable WNM filaments,
their typical properties are:
$n\sim50$ cm$^{-3}$, $T\sim80$ K, $R\sim0.1$ pc. 
The number of cold clouds, as well as their properties,  depend heavily
on the properties of the underlying turbulent flows. 

{\it General ISM turbulence} 
envisions interstellar clouds as dynamic entities that are constantly
changing in response to the turbulent ``weather'', a picture 
very different from the traditional approach. 
Numerical simulations of turbulent ISM, for example
Vazquez-Semadeni, Ballesteros-Paredes, \& Rodriguez (1997) show a lot of clouds
with very small sizes and low column densities ($<10^{19}$ cm$^{-2}$).
These clouds are out of dynamical equilibrium and probably very transient. 

{\it CNM destruction by shocks} in simulations by Nakamura et al. (2005)
can also produce a spray of small HI ‘shreds’ that could be related to
low-$N$(HI) clouds.  These shreds have large aspect ratios, up to 2000, but
further comparison with observations awaits inclusion of magnetic field.

\section{Summary}
The CNM clouds with $N({\rm HI})\sim 10^{18}$ cm$^{-2}$ are 
easily and frequently detected with deep radio integrations. 
These clouds are very thin along the line of sight, $L(||)\sim800$--4000 AU,
and could be related to TSAS.
They are evaporating very fast, unless being surrounded by a lot 
of mild WNM, $T\sim10^{2} - 10^{4}$ K, in which case could last 
for up to $\sim1$ Myr.
The CNM fraction relative to the total HI column density also supports the  
existence of large WNM ‘envelopes’ which contribute up to 95-99\% 
of the total $N$(HI).
The combined histogram of $N({\rm HI})$ for
`classic' CNM clouds and low-$N$(HI) clouds can be fitted with a single
function, $\Phi(N) \propto N^{-1}$.  
This suggests that low-$N$(HI) clouds are, most likely, a low
column density extension of the CNM cloud population.

\acknowledgements %%% Text of acknowledgements runs on after this command.
SS and CH acknowledge support by NSF grants AST-0097417 and 
AST-9981308. We would also like to thank the NRAO 
for hosting this highly stimulating and enjoyable conference.

%%% THE BIBLIOGRAPHY
%%%
%%% CONSULT SECTION 3 OF "INSTRUCTIONS FOR AUTHORS" FOR HOW TO USE NATBIB.
%%% AUTHORS ARE ENCOURAGED TO USE EITHER THE "THEBIBLIOGRAPY" ENVIRONMENT
%%% BY UNCOMMENTING (DELETING THE "%" SYMBOL) THE COMMANDS BELOW, OR BY
%%% USING THE BIBTEX ENVIRONMENT. TO FIND OUT WHICH IS APPLICABLE TO YOUR
%%% CONTRIBUTION, CONSULT THE VOLUME EDITORS FOR YOUR PROCEEDINGS.
%%%

\end{document}